# Generation of coupled orbital angular momentum modes from an optical vortex parametric laser source

ROUKUYA MAMUTI,[1] SHUNSUKE GOTO,[1] KATSUHIKO MIYAMOTO,[1,2] AND TAKASHIGE OMATSU,[1,2,*]

[1] *Graduate School of Advanced Integration Science, Chiba University, 1-33 Yayoi-cho, Inage-ku, Chiba 263-8522, Japan*
[2] *Molecular Chirality Research Center, Chiba University, 1-33 Yayoi-cho, Inage-ku, Chiba 263-8522, Japan*
*omatsu@faculty.chiba-u.jp

**Abstract:** We report on the generation of flower (wheel) modes, which manifest coupled orbital angular momentum (OAM) modes, from a vortex pumped optical parametric oscillator simply by employing a pump source with a short temporal coherence time. This vortex oscillator was also developed to generate a further higher-order vortex signal output with $\ell_s$=2–4 by replacement of the pump source with a longer coherence time. The signal and idler outputs were tuned within wavelength ranges of 745–955 nm and 1200–1855 nm, respectively. The maximum signal output energy of 1.2 mJ was measured with an optical efficiency of 15.6%.

## 1. Introduction

An optical vortex exhibits a ring-shaped spatial intensity profile and an orbital angular momentum (OAM) of $\ell\hbar$ [1–5] arising from its azimuthal phase term $e^{i\ell\phi}$, where $\ell$ is an integer referred to as the topological charge. The handedness of the optical vortex is also determined by the sign of the topological charge. The optical vortex has been widely applied in a variety of fields, such as optical trapping and manipulation [6–12], optical metrology [13–15], fiber-based or free-space optical communication [16–24], quantum computing [25–27], cold atom physics [28–32], astrophysics [33–35], and material processing [36,37]. Furthermore, an optical vortex enables the twisting of a variety of materials, such as metals [38–41], semiconductors [42,43], azopolymers [44–46], and even liquid-resin to shape helical nano/microscale structures. Such helical structures are expected to offer new research avenues, for example, chiral selective imaging systems (e.g., atomic force microscopes [47]), optoelectro-mechanical systems, and plasmon-enhanced chiral metamaterials.

Frequency extension of the optical vortex via nonlinear optical interactions [48], such as second harmonic generation [49–55], sum frequency generation [56,57], optical parametric generation [58–60], and stimulated Raman scattering [61] provides not only wavelength versatile optical vortex sources that match their wavelengths with the absorption bands of individual materials for such applications but also a variety of new fundamental physical insights. In particular, an optical parametric oscillator (OPO) provides an efficient method to achieve widely tunable optical vortex sources in the near- and mid-infrared regions, and it also inspires a question concerning the OAM conservation law, that is, how the OAM of a pump beam is divided between the signal (higher energy photon) and idler (lower energy photon) outputs [62].

To date, we have successfully demonstrated a nanosecond widely-tunable vortex source formed of a singly resonant non-critical phase-matching LiB$_3$O$_5$ (LBO) OPO. In this system, the generated signal and idler outputs exchange their OAMs simply by appropriately shortening

(or extending) the cavity [62]. Furthermore, the signal output is transformed into three different OAM modes with topological charges of $\ell_s$=0–2 by only tuning their lasing wavelengths [63]. Here, we reveal that this system enables the creation of various flower (wheel) modes (termed coupled OAM modes), which manifest a coherent superposition of two OAM modes as a signal (idler) output, simply by employing a pump source with a short temporal coherence length. We also address that this vortex source generates a further higher-order vortex signal output with $\ell_s$=2–4 by replacement of the pump source.

## 2. Experiments

A schematic diagram of the experimental setup is shown in Fig. 1(a). The pump source used was a frequency-doubled, diode-pumped, Q-switched Nd:YAG laser (pulse width: ca. 10 ns, wavelength: 532 nm, beam propagation factor, $M^2$: ca. 1.1), the output of which was converted into an optical vortex with a topological charge $\ell_p$ of 2 using a spiral phase plate [64] (RPC Photonics, VPP-1c).

A laser cavity was formed from a flat input mirror with high reflectivity at 0.65–1.05 µm and a concave (curvature radius: 500 mm) output mirror with 90% reflectivity at 0.8 µm, to form a singly resonant cavity for the signal output. A 45 mm non-critical phase-matching LBO crystal ($\theta$=90°, $\varphi$=0°) was used as a nonlinear medium and was mounted on an oven to control the temperature within the range of 128.5–182.4 °C. The facets of the crystal were also anti-reflection (AR)-coated for 532 nm and 1064 nm. The cavity length was measured to be ca. 60 mm. It should be noted that the pump source exhibited a coherence time of ca. 15 ps, ($\gamma$ is the visibility of the fringes, so called the coherence degree, measured by a conventional Michelson interferometric technique [65].) which is much shorter than the intracavity round trip time of photons (ca. 570 ps) (Fig. 1(b)). The generated signal and idler outputs were observed using a Si CCD camera and an InGaAs camera, respectively.

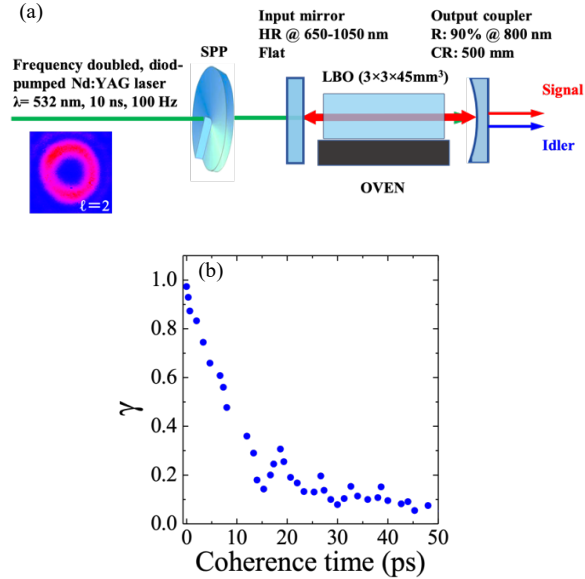

Fig. 1(a) Experimental setup of singly resonant optical parametric oscillator. (b) Measured temporal coherence of pump laser.

The signal output showed a flower-shaped signal mode with 4 petals (Fig. 2(a)). Such a flower signal mode $u(r, \phi)$ should manifest the coupled OAM modes, *i.e.*, the coherent superposition of two OAM modes with $\ell_s$ and $-\ell_s$, as given by the formula

$$u(r,\phi) \propto \alpha \cdot r^{|\ell_s|} e^{-\left(\frac{r}{\omega_0}\right)^2} e^{i\ell_s\phi} + \beta \cdot r^{|\ell_s|} e^{-\left(\frac{r}{\omega_0}\right)^2} e^{-i\ell_s\phi} \quad (1)$$

$$\alpha^2 + \beta^2 = 1 \quad (2)$$

where $r$ and $\phi$ are respectively the radial and azimuthal components in cylindrical coordinates, and $\omega_0$ is the beam waist of the OAM mode. This theoretical formula with $\ell_s=2$ supports the experimental data well (Fig. 2(g)). We then fixed $\alpha$ to be 0.32. The coupled OAM modes were also evidenced by the X-shaped fringes formed of the orthogonal Hermite-Gaussian $HG_{0,2}$ and $HG_{2,0}$ modes observed in the focused signal with the use of a cylindrical lens, as shown in Fig. 2(d) [58].

When the signal output was tuned towards the blue region, the system allowed the signal output to lase in further higher-order ($\pm 3^{th}$ or $\pm 4^{th}$) coupled OAM modes with a larger dark core, thereby yielding flower modes with 6 or 8 petals (Figs. 2(b), (c) and (e), (f)).

A maximum 930 nm signal output energy of 1.18 mJ was measured at a pump energy of 7.7 mJ, which corresponds to an optical efficiency of 15.3%. The signal mode size $\omega_s$ in the system is almost proportional to the signal wavelength as reported in our previous publication [63]. Thus, the signal output is allowed to lase in further higher-order ($\pm 3^{rd}$ or $\pm 4^{th}$) coupled OAM modes with 6 or 8 petals, as the wavelength shifts toward the blue region. These experiments are also supported by Eq. (3) (Figs. 2(h) and (i)).

Beyond the scope and motivation of this work, further experimental investigation and theoretical analysis, including gain dynamics, spatial (and longitudinal) mode competition, as well as thermal issues of the crystal will be necessary to explain such the mechanism of the coherent coupling (phase-locking) between two LG modes with the positive and negative charges.

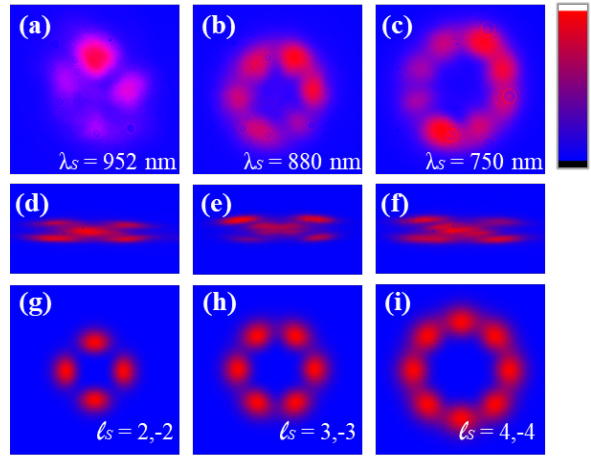

Fig. 2 (a–c) Flower modes with $\ell_s=\pm 2, \pm 3$, and $\pm 4$. (d–f) Wavefront measurement of signal outputs with a tilted lens. (g–i) Theoretical flower modes with $\ell_s =\pm 2, \pm 3$ and $\pm 4$. $\alpha$ is 0.32.

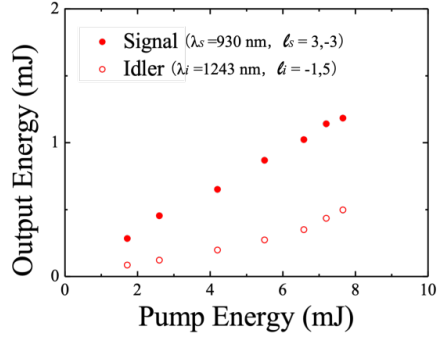

Fig. 3 Power scaling of the 930 nm signal and corresponding 1243 nm idler outputs.

According to the OAM conservation among the pump, signal and idler outputs, the corresponding idler output exhibited a wheel-shaped mode, which manifests the coherent superposition of two OAM modes with $2-\ell_s$ and $2+\ell_s$ given by the expression

$$u(r,\phi) \propto \alpha \cdot r^{|2-\ell_s|} e^{-\left(\frac{r}{\omega_0}\right)^2} e^{i(2-\ell_s)\phi} + \beta \cdot r^{|\ell_s+2|} e^{-\left(\frac{r}{\omega_0}\right)^2} e^{-i(\ell_s+2)\phi} \qquad (4)$$

The theoretical formula ($\alpha$= 0.32) in Eq. (4) supports the experimental results, as shown in Fig. 4. It should be noted that the 1243 nm idler output energy was then measured to be 0.49 mJ with an optical efficiency of 6.4%.

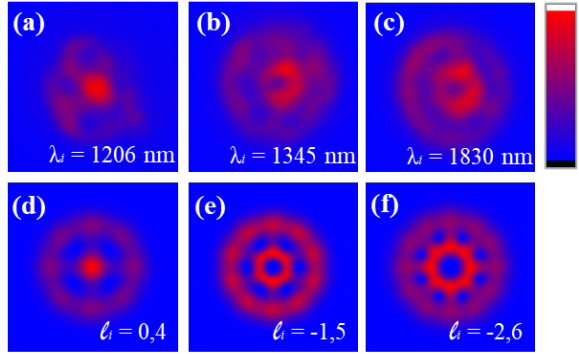

Fig. 4(a–c) Wheel-shaped idler modes with $\ell_i$=0 & 4, -1 & 5, and -2 & 6. (d–f) Theoretical wheel-shaped idler outputs.

## 3. Discussion

A pump source with a rather short coherence time (ca. 15 ps) should allow the cavity to operate under the condition of near optical parametric generation, thereby enabling the creation of the coupled OAM modes. The experiments were also conducted using another 532 nm pulsed laser with a pulse duration of ca. 10 ns and a long coherence time (see Fig. 5) as a pump source. The cavity configuration (such as the nonlinear crystal, length, input mirror, and output coupler) was identical with that in the experiments performed by the other nanosecond pulsed laser with the short coherence time.

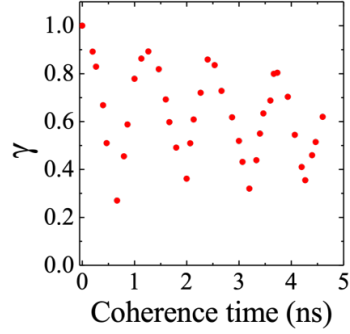

Fig. 5 Coherence of pump pulse

The signal output had a donut-shaped spatial form as a vortex mode with topological charges of $\ell_s$=4, 3, and 2, as evidenced by the tilted $HG_{0,\ell_s}$ mode with $|\ell_s|+1$ as bright lobes converted by the cylindrical lens [66,67]. The wavelength of the signal output was then in the range of 745–955 nm.

The nonlinear parametric gain in the OPO can generally be determined by the spatial overlap efficiency $\eta_{\ell p,\ell s,\ell i}$ among the pump and resonating signal amplitudes, as noted in the previous publications [68],

$$\eta_{\ell_p,\ell_s,\ell_i} = \left|\int E_p E_s^* E_i^* \, dS\right| \propto \int_0^{2\pi}\int_0^\infty r^{|\ell_p|+|\ell_s|+|\ell_i|} \exp\left(-\left(\frac{1}{\omega_p^2}+\frac{1}{\omega_s^2}+\frac{1}{\omega_i^2}\right)r^2\right) \cdot \exp(i(\ell_p - \ell_s - \ell_i)\phi) \, r \, dr d\phi \quad (3)$$

where $r$ is the radial coordinate, and $\omega_p$, $\omega_s$, and $\omega_i$ are the mode sizes of the pump, signal, and idler, respectively. The signal output should be allowed to lase at the vortex mode with maximum $\eta_{\ell p,\ell s,\ell i}$ as a cavity mode by employing a pump source with a long coherent time. Our system thus enables us to generate the vortex signal output only with a positive topological charge $\ell_s$ as a cavity mode. The resulting idler output then exhibits the vortex mode with the topological charge of $\ell_p-\ell_s$. For instance, under a nearly degenerate condition (the signal wavelength ~953nm), our system enables us to generate the vortex signal output with a topological charge $\ell_s$=2. The resulting idler output then exhibits Gaussian spatial form without any OAM. In fact, $\eta_{2,2,0}$ is approximately 15 times larger than $\eta_{2,-2,4}$.

However, when the pump source exhibits a shorter coherence time, namely in the case of optical parametric generation and amplification, the system should allow the generation of the signal and idler outputs with diverse OAM states. In particular, two signal vortex outputs with positive and negative topological charges exhibit the same spatial overlap of intensity profile with the pump, thereby generating efficiently both vortex modes, so as to create the flower modes.

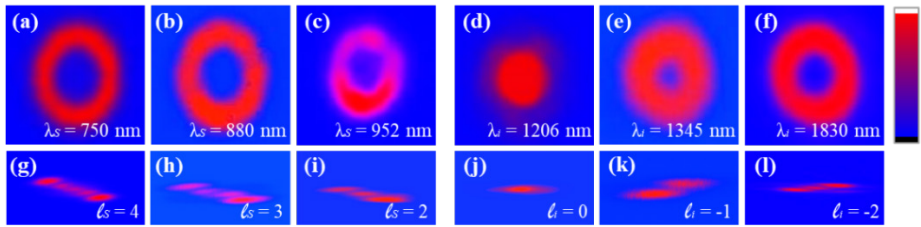

Fig. 6 Spatial profiles of (a–c) signal (750 nm, 880 nm, and 952 nm) and (d–f) idler (1830 nm, 1345 nm, and 1206 nm) vortex outputs. Inclined focused spatial forms of (g–i) signal (770 nm, 880 nm, and 952 nm) and (j–l) idler (1830 nm, 1345 nm, and 1206 nm) outputs.

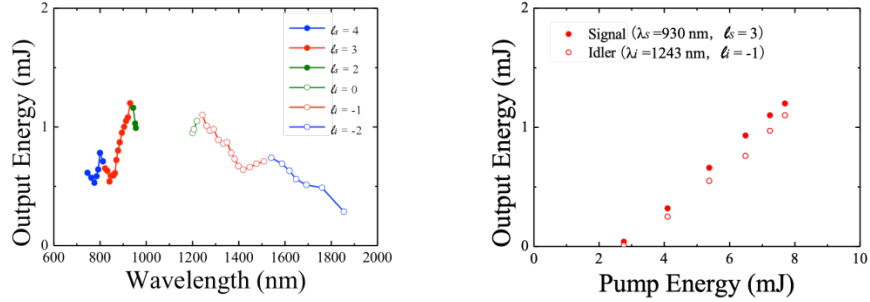

Fig. 7. Tunability and power scaling of signal and idler outputs.

The resulting idler output showed the vortex mode with topological charges of $\ell_i$=0, -1, and -2, so that the OAM among the pump, signal and idler outputs was conserved. The wavelength was tuned in the range of 1200–1855 nm (Fig. 7). The maximum signal output energy of 1.2 mJ was then observed at a signal wavelength of 930 nm with a pump energy of 7.7 mJ for the signal vortex output, which corresponds to an optical efficiency of 15.6%. The corresponding idler output energy was measured to be 1.1 mJ.

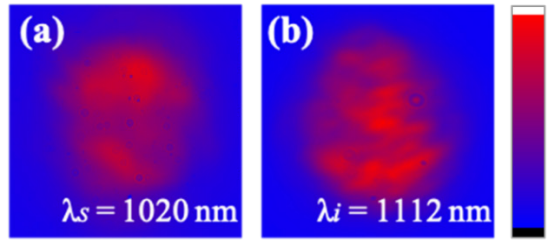

Fig. 8(a) 1020 nm signal and (b) corresponding 1112 nm idler outputs from the double resonance region.

It should be noted that the mixed vortex mode output, formed of incoherently coupled vortex and Gaussian modes, was generated within the wavelength region of 955–1200 nm, due to the double resonance of the signal and idler outputs (Fig. 8).

## 4. Conclusion

We have successfully demonstrated, for the first time to our best of knowledge, the generation of flower (wheel) modes (coupled OAM modes), formed by the coherent superposition of two OAM modes from a vortex-pumped OPO, simply by employing a pump source with a short temporal coherence length.

We also address that this vortex source generates a further higher-order vortex signal output with $\ell_s$=2–4 by employing a pump source with a long temporal coherence length. The signal and idler outputs were tuned within the wavelength ranges of 745–955 nm and 1200–1855 nm, respectively. The maximum signal output energy of 1.2 mJ was measured with an optical efficiency of 15.6%.

Even an entangled state of the signal and idler, in general, manifests an incoherent superposition of modes [69]. Such system with a flower or wheel mode will offer us a new physical insight and a significant advance in optical science and photonics, and it will also be potentially applied in a variety of fields, such as for optical trapping and manipulation. The system will further be extended to generate flower or wheel modes in the mid-infrared or terahertz regions. Further OAM versatility in this system will be realized through the use of different cavity configurations, such as a singly resonant cavity for an idler.

## Supplements

It is worth noting that the signal output occasionally exhibits a ring-shaped mode formed of incoherent superposition (phase-unlocking) of the LG modes with $\ell_s$ and $-\ell_s$. The resulting idler then shows a double-ring spatial form, manifesting incoherent superposition (phase-unlocking) of the LG modes with $\ell_p-\ell_s$ and $\ell_p+\ell_s$. Such switching of coherent and incoherent coupling between two LG modes will be investigated as a future work.

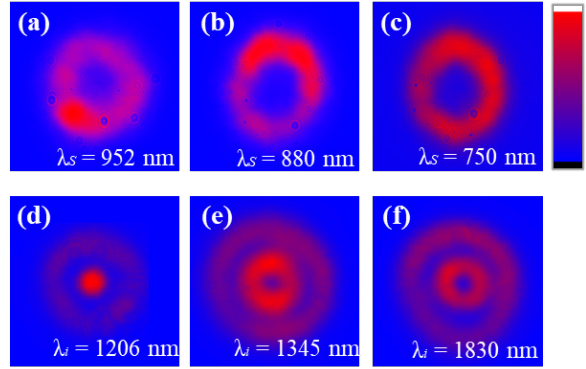

Fig. 9 (a–c) Ring-shaped modes with $\ell_s$=±2, ±3, and ±4. (d-f) Double-ring modes with $\ell_i$=-4 and 0, -3 and 1, and -2 and 6.

## Funding
KAKENHI, Japan Society for the Promotion of Science (JP16H06507, JP17K19070, and JP18H03884); CREST, Japan Science and Technology Agency (JPMJCR1903).

## Disclosures
The authors declare no conflicts of interest.